\documentclass[pra,aps,showpacs,floatfix,twocolumn,reprint,superscriptaddress]{revtex4-1}
\usepackage{amsmath}
\usepackage{amsfonts}
\usepackage{amssymb}
\usepackage{revsymb}
\usepackage{graphicx}
\usepackage{dcolumn}

\newcommand{\sref}[1]{Sec. \ref{#1}}
\newcommand{\eref}[1]{Eq.~(\ref{#1})}
\newcommand{\tref}[1]{Table~\ref{#1}}
\newcommand{\fref}[1]{Fig.~\ref{#1}}

\graphicspath{{figs/}}

\def\ket#1{|\,#1 \,\rangle}

\def\vop#1{\mathbf{\hat{#1}}}

\mathchardef\mhyphen="2D
\begin{document}
\title{Measurements of the branching ratios for $6P_{1/2}$ decays in $^{138}$Ba$^+$}
\author{K. J. Arnold}
\affiliation{Centre for Quantum Technologies, National University of Singapore, 3 Science Drive 2, 117543 Singapore}
\author{S. R. Chanu}
\affiliation{Centre for Quantum Technologies, National University of Singapore, 3 Science Drive 2, 117543 Singapore}
\author{R. Kaewuam}
\affiliation{Centre for Quantum Technologies, National University of Singapore, 3 Science Drive 2, 117543 Singapore}
\author{T. R. Tan}
\affiliation{Centre for Quantum Technologies, National University of Singapore, 3 Science Drive 2, 117543 Singapore}
\affiliation{Department of Physics, National University of Singapore, 2 Science Drive 3, 117551 Singapore}
\author{L. Yeo}
\affiliation{Department of Physics, National University of Singapore, 2 Science Drive 3, 117551 Singapore}
\author{Zhiqiang Zhang}
\affiliation{Centre for Quantum Technologies, National University of Singapore, 3 Science Drive 2, 117543 Singapore}
\author{M. S. Safronova}
\affiliation{Department of Physics and Astronomy, University of Delaware, Newark, Delaware 19716, USA}
\affiliation{Joint Quantum Institute, National Institute of Standards and Technology and the University of Maryland,
College Park, Maryland, 20742}
\author{M. D. Barrett}
\email{phybmd@nus.edu.sg}
\affiliation{Centre for Quantum Technologies, National University of Singapore, 3 Science Drive 2, 117543 Singapore}
\affiliation{Department of Physics, National University of Singapore, 2 Science Drive 3, 117551 Singapore}
\date{\today}

\begin{abstract}
Measurement of the branching ratios for $6P_{1/2}$ decays to $6S_{1/2}$ and $5D_{3/2}$ in $^{138}$Ba$^+$ are reported with the decay probability from $6P_{1/2}$ to $5D_{3/2}$ measured to be $p=0.268177\pm(37)_\mathrm{stat}-(20)_\mathrm{sys}$.  This result differs from a recent report by $12\sigma$.  A detailed account of systematics is given and the likely source of the discrepancy is identified.  The new value of the branching ratio is combined with a previous experimental results to give a new estimate of $\tau=7.855(10)\,\mathrm{ns}$ for the $6P_{1/2}$ lifetime.  In addition, ratios of matrix elements calculated from theory are combined with experimental results to provide improved theoretical estimates of the $6P_{3/2}$ lifetime and the associated matrix elements.
\end{abstract}
\maketitle
\section{Introduction}
Singly-ionized Barium has been well studied over the years with a wide range of precision measurements \cite{marx1998precise,knoll1996experimental,hoffman2013radio,woods2010dipole,sherman2008measurement,kurz2008measurement,munshi2015precision,dutta2016exacting} that have provided valuable benchmark comparisons for theory \cite{guet1991relativistic,dzuba2001calculations,iskrenova2008theoretical,safronova2010relativistic,sahoo2007theoretical,gopakumar2002electric}.  With the proposed parity nonconservation (PNC) measurement using the $S_{1/2}-D_{3/2}$ transition in $^{137}$Ba$^+$ \cite{fortson1993possibility},  there has been particular interest in decays from the $6P_{1/2}$ level.  This level has an estimated $\sim90\%$ contribution to the parity-violating electric dipole transition amplitude between the $S_{1/2}$ and $D_{3/2}$ states \cite{dzuba2001calculations}.  Consequently lifetime measurements of the $6P_{1/2}$ level and the associated branching fractions provide important benchmarks for the interpretation of a PNC experiment.  This was a motivating factor in the recent branching ratio measurements for the $6P_{1/2}$ level, which reported results with fractional inaccuracies of $0.03\%$ \cite{munshi2015precision}.

Our own interest in the branching ratio is of relevance to polarisability assessments in optical atomic clocks.  Measured atomic properties of $^{138}$Ba$^+$ can be combined to provide an accurate model of the dynamic differential scalar polarisability $\Delta\alpha_0(\omega)$ for the $S_{1/2}-D_{5/2}$ clock transition \cite{BarrettProposal}.  With such a model, ac-Stark shifts of the clock transition could provide an \emph{in situ} calibration of laser intensities. Properties of interest are the reduced electric dipole matrix elements $\langle{6 P_{1/2}}\|r\|{6 S_{1/2}}\rangle$, $\langle{6 P_{3/2}}\|r\|{6 S_{1/2}}\rangle$ and $\langle{6 P_{3/2}}\|r\|{5 D_{5/2}}\rangle$, which determines the dominate contributions to $\Delta\alpha_0(\omega)$, and the zero crossing of $\Delta\alpha_0(\omega)$ near 651\,nm, which determines an overall offset.  Consequently, we have sought to independently confirm relevant measurements from the literature, which includes a branching ratio measurement for the $6P_{1/2}$ level.  Measurements are carried out using a single ion in a similar manner as for $^{138}$Ba$^+$ \cite{munshi2015precision}, $^{40}$Ca$^+$ \cite{ramm2013precision}, $^{88}$Sr$^+$ \cite{likforman2016precision,zhang2016iterative}, and $^{226}$Ra$^+$ \cite{fan2019measurement}.  However the result differs by approximately $12\sigma$ from a previous report in the literature \cite{munshi2015precision}.

The report is divided into three main sections.  The first section outlines the experimental setup and details the results obtained.  The second section compares the result to theory.  Combining the measured branching ratio with the experimental results in \cite{woods2010dipole} provides an experimental determination for the $6P_{1/2}$ lifetime and the reduced dipole matrix element $\langle{6 P_{1/2}}\|r\|{5 D_{3/2}}\rangle$.  Matrix element ratios calculated from theory are combined with experimental results to provide improved estimates of the $6P_{3/2}$ lifetime and the associated matrix elements.  The third section gives an account of the systematic effects relevant to a $P_{1/2}$ branching ratio measurement. This includes a detailed discussion of detector imperfections, which we believe is the likely source of the discrepancy with the result in \cite{munshi2015precision}.

\section{Experiment}

\begin{figure}
\includegraphics[width=\columnwidth]{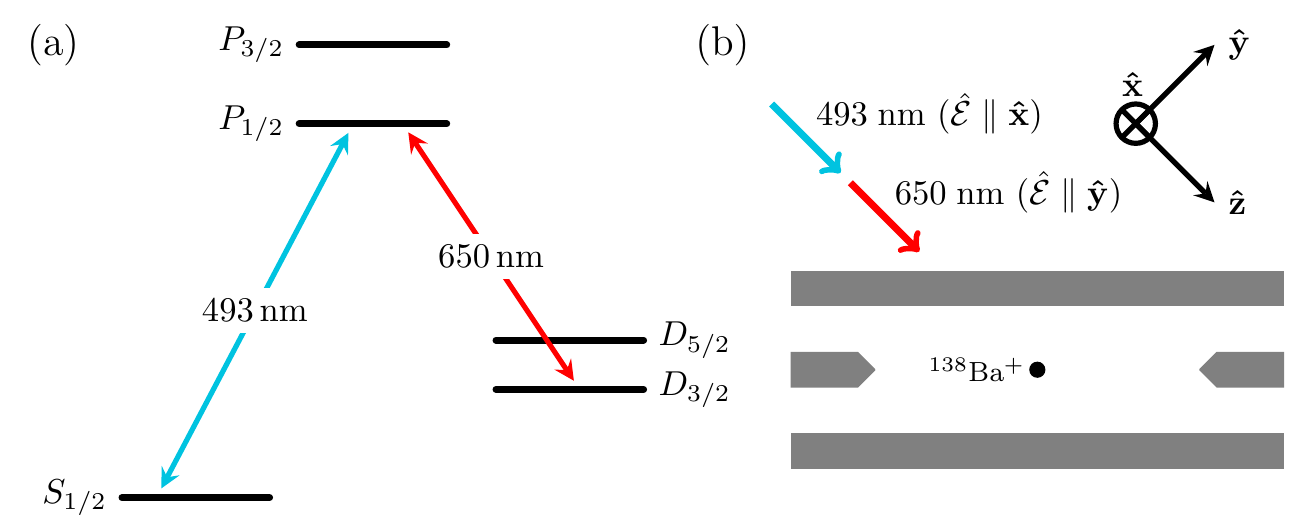}
\caption{(a) Low-lying level structure of $^{138}$Ba$^+$ showing transitions relevant to this work. (b) Schematic of experiment geometry. The emission from a single Ba$^+$ is imaged along the $\vop{x}$ axis onto a single photon counting module (SPCM).}
\label{fig:levels}
\end{figure}

The experiments are carried out in a linear Paul trap with axial endcaps, the details of which have been given elsewhere \cite{arnold2018blackbody,kaewuam2018laser}.  The relevant level structure of $^{138}$Ba$^+$ is illustrated in \fref{fig:levels}(a), which shows the the two transitions $S_{1/2}-P_{1/2}$ and $D_{3/2}-P_{1/2}$ with resonant wavelengths near 493\,nm and 650\,nm respectively.  As illustrated in  \fref{fig:levels}(b), the beams copropagate orthogonal to the imaging axis ($\vop{x}$) and at 45 degrees to the trap axis.  Both beams are linearly polarized with the 493-nm (650-nm) laser polarized along (perpendicular to) the imaging axis.  In all experiments this configuration is unchanged.  For reference purposes a coordinate system is defined by the imaging axis ($\vop{x}$), 650-nm polarization ($\vop{y}$), and beam propagation direction ($\vop{z}$).  An applied magnetic field typically within the range of 0.1 to 0.4\,mT is sufficient to lift the Zeeman degeneracy and prevent dark states when driving the $D_{3/2}-P_{1/2}$ transition with 650\,nm light.  As discussed later in the section multiple orientations of the magnetic field have been used.

We are restricted to collecting 650-nm fluorescence due to the dielectric coating on the vacuum window.  The fluoresence from the ion is collected using an off-the-shelf aspheric lens with a specified numerical aperture of 0.42, and imaged through a 650-nm narrow-band filter onto a single photon counting module (SPCM) with specified quantum efficiency of 65\%. Our observed  detection efficiency of $q\approx2.7\%$ is within 10\% of these specifications.

A $P_{1/2}$ branching ratio measurement is, in principle, a very straightforward experiment.  When optically pumping from $S_{1/2}$ to $D_{3/2}$ with the 493-nm beam, the atom scatters precisely one photon at 650\,nm.  The mean number of 650-nm photons detected is then a measure of the detection efficiency, $q$.  When optically pumping from $D_{3/2}$ to $S_{1/2}$ with the 650-nm beam, there is, on average, $p/(1-p)$ photons scattered at 650\,nm, where $p$ is the probability of decay from $P_{1/2}$ to $D_{3/2}$.  The mean number of 650-nm photons detected is then $r=pq/(1-p)$. With estimates of $r$ and $q$ from the average of $N$ measurement cycles, the branching ratio is then estimated from
\begin{equation}
\label{branchEq}
p=\frac{r}{r+q},
\end{equation}
which is independent of the detection efficiency.

\begin{figure}
\includegraphics[width=\columnwidth]{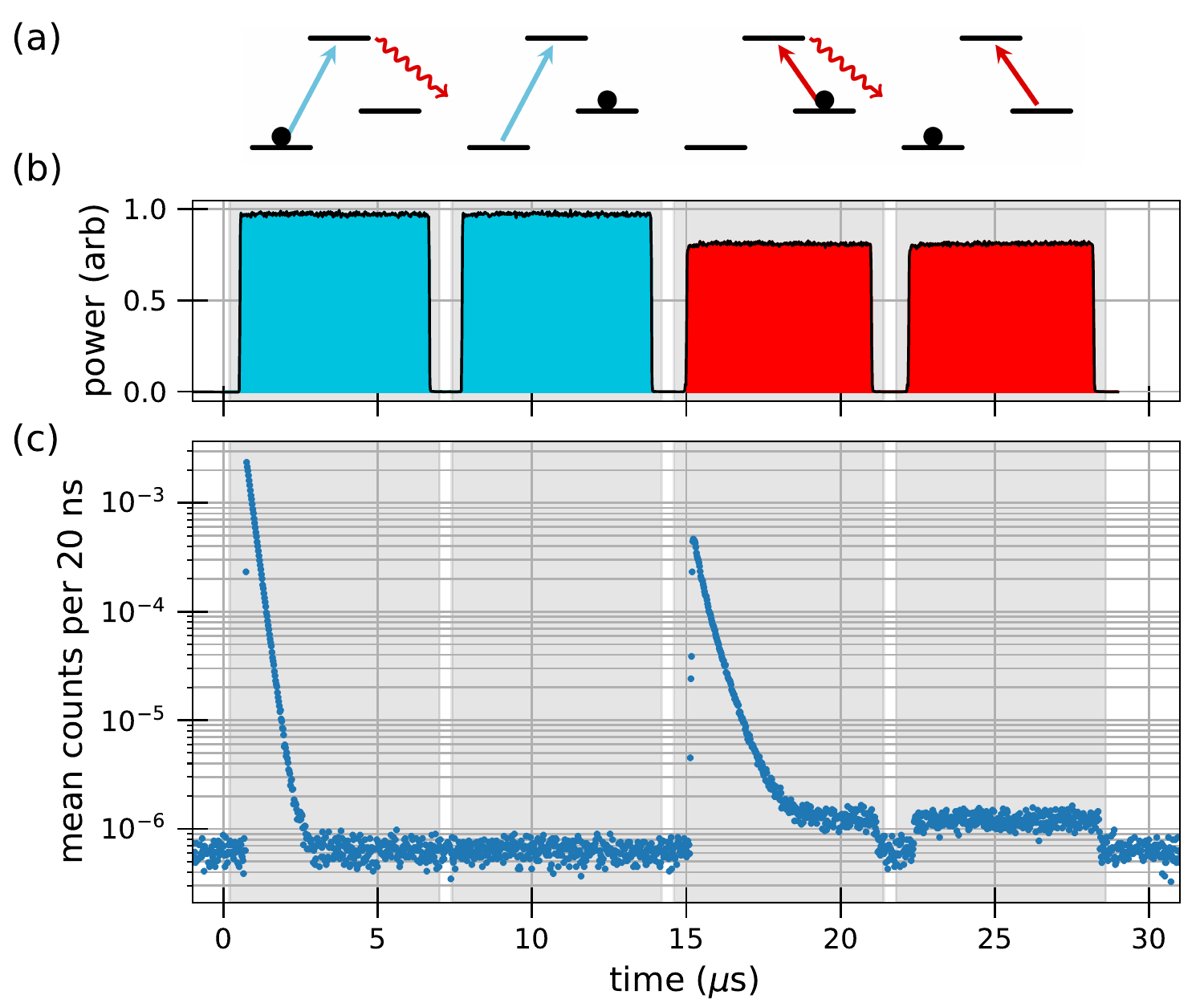}
\caption{(a) Sequence of pumping pulses. (b) Measured intensity profiles for 493 nm (aqua) and 650 nm (red) laser pulses. (c) Example of measured signals from $4.9\times10^7$ experiments using high resolution time-tagging.  For long data runs, all detection events within the grey shaded regions are accumulated to obtain the four respective averages.}
\label{fig:timing}
\end{figure}

 As illustrated in \fref{fig:timing}, a detection cycle consists of four pulses of equal duration $\tau$, which is typically $6\,\mathrm{\mu s}$.  The first two pulses are from the 493-nm laser with the first optically pumping the atom to $\ket{D_{3/2}}$ to measure $q$ and the second provides a measurement of the background contribution.  This is followed by a similar pair of pulses from the 650-nm beam to measure the signal and background contributions when pumping from $D_{3/2}$ to $S_{1/2}$.  With small delays between each pulse, the net cycle time is around $31\,\mathrm{\mu s}$.  

\begin{figure}
\includegraphics[width=\columnwidth]{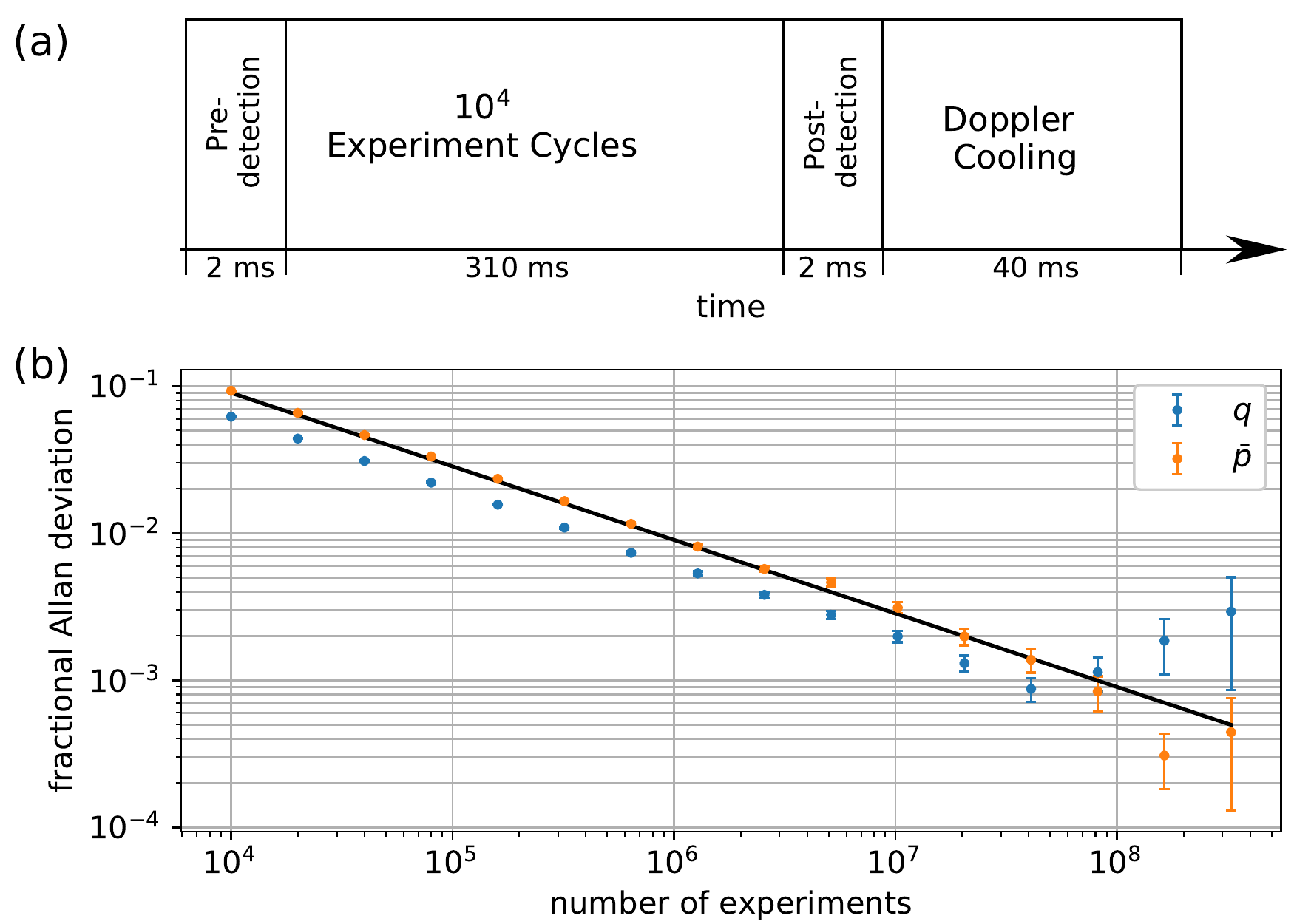}
\caption{(a) Timing sequence for data collection.  (b) Fractional Allan deviation of the measured detection efficiency, $q$, and branching ratio, $p$, in a typical dataset. The black line indicates the statistically limited uncertainty for $p$.}
\label{fig:qpallan}
\end{figure}

 Measurements are carried out in blocks of $10^4$ detection cycles.  As illustrated in \fref{fig:qpallan}(a),  between each block there is approximately 44 ms of Doppler cooling where both 493- and 650-nm beams are on.   For the 2 ms of Doppler cooling both immediately before and after each block, the number of 650-nm photons detected is recorded. This enables detection of both collisions and rare, off-resonant scattering to $D_{5/2}$.  Such events would compromise data integrity but can be identified by a statistically significant drop in the fluorescence collected during Doppler cooling.  When low counts are detected in either the pre- or post- detection step, that data block is discarded.  This typically results in approximately 0.35\% of the data being discarded, most of which is attributed to shelving to $D_{5/2}$. Only $\sim1.1\%$ of discarded blocks are attributed to collisions. Even though excitation to $D_{5/2}$ is somewhat less frequent, it results in an extended period ($\sim 30\,\mathrm{s}$) out of the cooling cycle. Figure~\ref{fig:qpallan}(b) shows the Allan deviation of the measured parameters $q$ and $p$ for a typical run of $\sim10^9$ experiments.  The detection efficiency $q$ is susceptible to thermal drift in the collection optics and we observe instability at the $10^{-3}$ level for long averaging times. However, the inferred branching ratio, $p$, is independent of $q$ and averages in accordance with the statistical limit (black line \fref{fig:qpallan}(b)).

\begin{figure}
\includegraphics[width=\columnwidth]{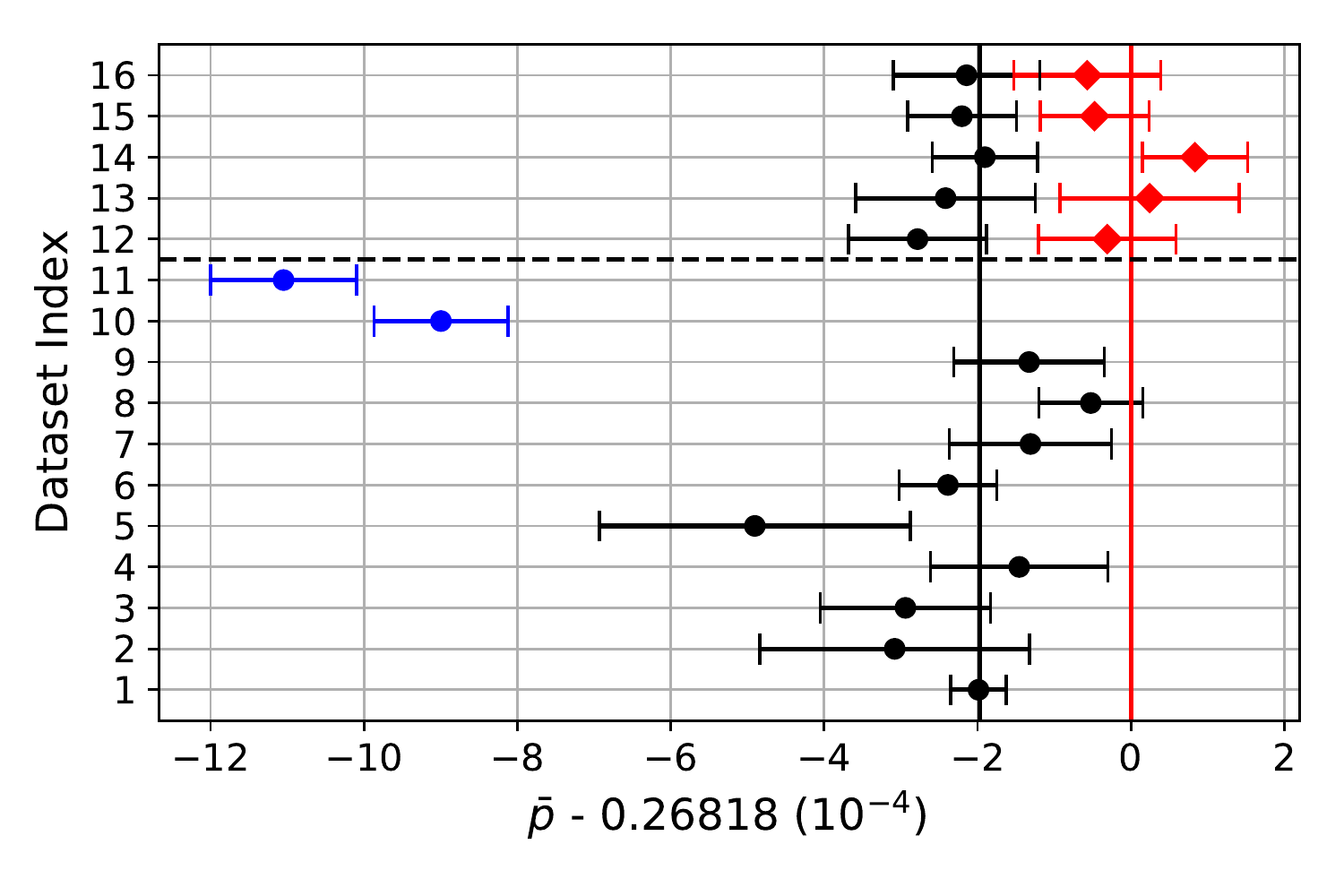}
\caption{(circles) Raw measurements of the branching ratio and the weighted mean (black line), excluding the two outliers as discussed in the text. The final five measurements are corrected for systematic effects (red diamonds) related to the detector deadtime and the background subtraction. The weighted mean of these corrected values (red line) is $p=0.268177$.}
\label{fig:results}
\end{figure}

Raw data from $\sim150$ hours of data acquisition taken over a $6$ month period is shown in \fref{fig:results} (circles) with the error bars representing the statistical error.  Optical pumping times with 493- and 650-nm light were measured for each dataset and spanned the range $140-310\,\mathrm{ns}$.  Each data set was taken under one of four different magnetic field orientations: $\vop{x}, \vop{z}, \vop{z}'$, and $\cos\theta \vop{x}+\sin\theta \vop{z}'$, where $\theta\approx 33^\circ$ and $\vop{z}'=(\vop{y}+\vop{z})/\sqrt{2}$ is the trap axis.  Excluding the two outlier points, the weighted mean value is $p=0.267979(21)$ with a reduced chi-square $\chi^2_\nu=0.80$ and represents a total of $\sim14$ billion experiments.  The uncertainty is the usual standard error in the weighted mean, and is within $12\%$ of the statistical error expected from the total number of experiments included. This suggests that systematic shifts are at least uniform at the level of the statistical errors over the range of conditions explored. 

For the experiments reported here, there are only two systematics that have a statistically significant contribution to the measured $p$. One arises from detector dead time, which lowers count rates dependent on the photon counting statistics; the other arises from an imbalance in the two pulses used to measure signal and background when optically pumping from $D_{3/2}$ to $S_{1/2}$. This results in an imperfect background subtraction resulting in a shift that is dependent on the signal-to-noise ratio (SNR).  A detailed discussion of these two systematics is given in Sect.~\ref{systematics}.  

The systematic arising from background subtraction was only discovered due to the occurrence of the two outliers (blue circles, \fref{fig:results}), which were taken on two consecutive days.  Consequently this systematic was only rigorously assessed for the last five data sets and only these data sets are used in the final analysis.  In \tref{tab:shifts}, the estimated shifts from dead time and background subtraction are tabulated for each of the five data sets, and the corrected estimates of $p$ shown in \fref{fig:results} (red diamonds).  A value of $p=0.268177(37)$ is then obtained from the weighted mean of the corrected estimates. Note that, if background corrections are applied to the earlier data using the calibration for the last five points, the weighted mean of all corrected data is $p=0.268197(21)$ with $\chi^2_\nu=0.71$, which is in agreement with the restricted data set. Nevertheless we report a final result of 
\[p=0.268177\pm(37)_\mathrm{stat}-(20)_\mathrm{sys},\] 
where the systematic uncertainty results from an uncalibrated dead-time effect and beam switching transients as discussed in Sec.~\ref{Deadtime:Measure} and \ref{Deadtime:Correction}.

\begin{table}
\label{tab:shifts}
\caption{Systematic shifts assessed for last five measurements.}
\begin{tabular}{c c c c c}
\hline
index & $\bar{p}$ & Background  & Deadtime  & $p$ \\\hline
12&0.267898(90) & 8.5[-5] & 1.62[-4] & 0.268146 \\
13&0.26793(12) & 9.9[-5]  & 1.68[-4] & 0.268201\\
14&0.267986(69) & 1.05[-4] & 1.69[-4] &0.268260\\
15&0.267956(71) & 1.09[-4] & 6.5[-5] &0.268129\\
16& 0.267962(96) & 5.8[-5]  & 9.9[-5]  &0.268120\\\hline
\end{tabular}
\end{table}

\section{Lifetime of $6P_{1/2}$ and comparison with theory}
The measured branching ratio together with the reduced electric-dipole matrix element $\langle{6 P_{1/2}}\|r\|{6 S_{1/2}}\rangle=3.3251(21)\,\mathrm{a.u.}$ reported in \cite{woods2010dipole} allows an experimental determination of the matrix element $\langle{6 P_{1/2}}\|r\|{5 D_{3/2}}\rangle$ and the $6P_{1/2}$ lifetime, for which we obtain $3.0413(21)\,\mathrm{a.u.}$ and $7.855(10)\,\mathrm{ns}$ respectively.  These can be compared to the theory values given in table~\ref{tab1}.

\begin{table*}[!htbp]
\caption{\label{tab1} Absolute values of reduced matrix elements (in a.u.), ratios $R_k$ as defined in the text, branching ratio $p$, and the $6P_{1/2}$ lifetime in ns. The all-order single-double (SD) and single-double + partial triple (SDpT) results are listed in SD and SDpT columns, corresponding scaled vales are listed in the SD$_{\text{sc}}$ and SDpT$_{\text{sc}}$ columns. Recommended values in the final column are based on experimental values where available: the first two matrix elements and $R_0$ are taken from \cite{woods2010dipole}, from these $\langle{6 P_{1/2}}\|r\|{5 D_{3/2}}\rangle$ and $\tau(6P_{1/2})$ are deduced from the results of this work, and the remaining two matrix elements and $\tau(6P_{3/2})$ are further inferred from theoretical estimates of $R_1$ and $R_2$ as described in the text.}
\begin{ruledtabular}
\begin{tabular}{lccccc}
\multicolumn{1}{c}{Properties} &
\multicolumn{1}{c}{SD} &
\multicolumn{1}{c}{SDpT} &
\multicolumn{1}{c}{SD$_{\text{sc}}$} &
\multicolumn{1}{c}{SDpT$_{\text{sc}}$}&
\multicolumn{1}{c}{Recommended} \\
\hline
$\langle{6 P_{1/2}}\|r\|{6 S_{1/2}}\rangle$      &     3.338  &  3.371    &  3.358     &   3.358  & 3.3251(21)\\
$\langle{6 P_{3/2}}\|r\|{6 S_{1/2}}\rangle$      &     4.710  &  4.757    &  4.738     &   4.738  & 4.7017(27)\\
$\langle{6 P_{1/2}}\|r\|{5 D_{3/2}}\rangle$&     3.050  &  3.096    &  3.076     &   3.065  & 3.0413(21) \\
$\langle{6 P_{3/2}}\|r\|{5 D_{3/2}}\rangle$ &    1.333	&1.353  &	1.344&	1.339 & 1.3285(13)\\
$\langle{6 P_{3/2}}\|r\|{5 D_{5/2}}\rangle$&     4.103  & 4.163     & 4.137      &   4.122  & 4.0911(31)  \\
$R_0$                   &    1.4109  &  1.4111   &  1.4111    &   1.4110   & 1.4140(17) \\
$R_1$                   &    1.3452  &  1.3448   &  1.3452    &   1.3451  & 1.3452(4)  \\
$R_2$                  &  0.4368	  &0.4371&  	0.4370 &	0.4369 &  0.4368(3)\\
 $p(\text{theory})$    &   0.2678  &  0.2698   &  0.2687    &    0.2673  & $0.268177 (57)$  \\
 $\tau(6P_{1/2})$      &   7.798  &  7.626   	&7.696	    &  7.711 & 7.855(10)\\
  $\tau(6P_{3/2})$      &   6.245  &  6.107   	&6.164	    &  6.175 & 6.271(8)
\end{tabular}
\end{ruledtabular}
\end{table*}

Matrix elements given in table~\ref{tab1} were calculated using the linearized coupled-cluster approach including single-double excitations (SD) and additional partial triple contributions (SDpT) as in Ref.~\cite{safronova2010relativistic}.  To estimate uncertainties in the theoretical results, two additional calculations, labeled SD$_{\text{sc}}$ and SDpT$_{\text{sc}}$, were carried out in which higher excitations are estimated using a scaling procedure as described in \cite{safronova2008all}.  The branching ratio, $p$, and lifetime of the $P_{1/2}$ level are calculated using the relevant matrix elements and experimental energies.  Agreement between the experimentally determined matrix elements and theoretical values is at the 1-2\% level and the maximum discrepancy in the lifetime is 3\%.  However, we note that the values in the column labelled SD give consistently excellent results; $\lesssim 0.5\%$ in the matrix elements and $\lesssim 1\%$ in the lifetime determination.

As is evident from the tabulated values, the branching ratio calculated via the different methods has a much smaller spread relative to the associated matrix elements.  Thus it would appear that $\langle{6 P_{1/2}}\|r\|{6 S_{1/2}}\rangle$ and $\langle{6 P_{1/2}}\|r\|{5 D_{3/2}}\rangle$ have similar correlation corrections which largely cancel in the ratio. This results in an excellent agreement between the theoretical values and that obtained in the experiment, with theoretical values having at most a $0.6\%$ difference with the experimental value.  

The ratio of matrix elements
\begin{subequations} 
\begin{align}
R_0&=\frac{\langle{6 P_{3/2}}\|r\|{5 S_{1/2}}\rangle}{\langle{6 P_{1/2}}\|r\|{5 S_{1/2}}\rangle},\\
R_1&=\frac{\langle{6 P_{3/2}}\|r\|{5 D_{5/2}}\rangle}{\langle{6 P_{1/2}}\|r\|{5 D_{3/2}}\rangle},\\
\intertext{and}
R_2&=\frac{\langle{6 P_{3/2}}\|r\|{5 D_{3/2}}\rangle}{\langle{6 P_{1/2}}\|r\|{5 D_{3/2}}\rangle},
\end{align}
\end{subequations}
can be calculated to a very high precision as they depend only weakly on correlation corrections.  Taking values from the SD column in table~\ref{tab1} with the maximum discrepancy between the different methods as an uncertainty gives $R_1=1.3452(4)$ and $R_2=0.4368(3)$.  These values, combined with the experimental value of $\langle{6 P_{1/2}}\|r\|{5 D_{3/2}}\rangle=3.0413(21)\,\mathrm{a.u.}$, gives new recommended values for $\langle{6 P_{3/2}}\|r\|{5 D_{5/2}}\rangle$ and $\langle{6 P_{3/2}}\|r\|{5 D_{3/2}}\rangle$ of $4.0911(31)\,\mathrm{a.u.}$ and $1.3285(13)\,\mathrm{a.u.}$, respectively.  Combined with the matrix elements given in \cite{woods2010dipole}, we then obtain an estimate for the $P_{3/2}$ lifetime of $6.271(8)\,\mathrm{ns}$.  The uncertainty is dominated by uncertainties in the matrix elements from \cite{woods2010dipole}, and we have assumed these are maximally correlated.

It is also worth noting that branching ratios for decays from the $P_{3/2}$ level can be written entirely in terms of the ratios $R_k$, and $p$.  Consequently these can also be accurately estimated.  For decays to $S_{1/2}$, $D_{3/2}$ and $D_{5/2}$, we estimate probabilities of $0.7428(6)$, $0.0279(1)$, and $0.2293(6)$, respectively, which are in disagreement with \cite{dutta2016exacting} at the level of $\sim2-4$ times the reported measurement uncertainties.  Given that the experimental value of $R_0=1.4140(17)$ is also in disagreement with the theoretical estimate of 1.4109(2), it would be of interest to have an independent assessment of all measurements.
 
\section{Systematics in a $P_{1/2}$ branching ratio measurement.}
\label{systematics}
The systematics for a branching ratio measurement can be broadly categorised as fundamental, technical, or practical:  fundamental limitations are imposed by the properties of the atom, such as the finite lifetime of the $D_{3/2}$ level or off-resonant scattering to the $P_{3/2}$ state; technical limitations arise from experimental imperfections, such as pump beam switching or detector limitations; practical limitations arise from the finite duration of the experiment.   For this work, precision is  limited by the finite duration of the experiment and accuracy is mostly limited by detector dead time and switching imperfections.  In this section we give a detailed discussion of these two limiting systematics and how they have been assessed. We also include a discussion of concerns raised in previous work \cite{munshi2015precision,ramm2013precision,fan2019measurement} of polarization selectivity in the detection optics in Sect.~\ref{Detection}.  The choice of field orientations used here was, in part, motivated by these concerns.  All other systematics of which we are aware are below the statistical measurement precision and are briefly described in Sect.~\ref{Misc}.  

\subsection{Background subtraction}
\label{Background}
The two outlier points shown in \fref{fig:results} were taken on two consecutive days and prompted an investigation as to the cause of the discrepancy.  The two points had a higher background from the 650-nm beam resulting in a degraded signal-to-noise ratio (SNR). Additionally a prior change in the optics had altered the alignment of AOM switch further suggesting the shift was related to the background subtraction.  To test for this, a background check was done by running the experiment without an ion and it was found that the mean counts obtained from the two 650-nm pulses differed by 2.8(5)\% resulting in an incorrect background subtraction.  This effect, together with the lower SNR, largely explained the two outliers.

Subsequent realignment of the AOM switch reduced the difference in mean counts to $\sim 1\%$ but this is still enough to cause a statistically significant systematic shift for the typical SNR obtained in any given data set.  To quantify this better it is necessary to distinguish the 650-nm contribution to the background from other sources.  Running the usual experiment for 24 hours without an ion and the 650-nm laser blocked gave mean counts of $0.021780(39),\,0.021777(39),\,0.021741(39),$ and $0.021755(39)$ for the four signals.  This confirmed that the 493-nm beam does not significantly contribute to the background, which \fref{fig:timing}(d) also suggests. Thus it is reasonable to denote the background for the first and second 650-nm pulses by $r_{b1}=q_b+r_l$ and $r_{b2}=q_b+\alpha r_l$ respectively, where $q_b$ is the background as determined by the second 493-nm pulse.  Thus, without an ion, $\alpha$ can be measured via
\begin{equation}
\alpha=\frac{r_{b2}-q_b}{r_{b1}-q_b}.
\end{equation}
When measuring $\alpha$, the 650-nm background contribution is increased by removing a neutral density filter to improve the SNR.  With an ion, the background used to correct the first 650-nm pulse is then determined by $\alpha^{-1}(r_{b2}-q_b)+q_b$.

\begin{figure}
\includegraphics[width=\columnwidth]{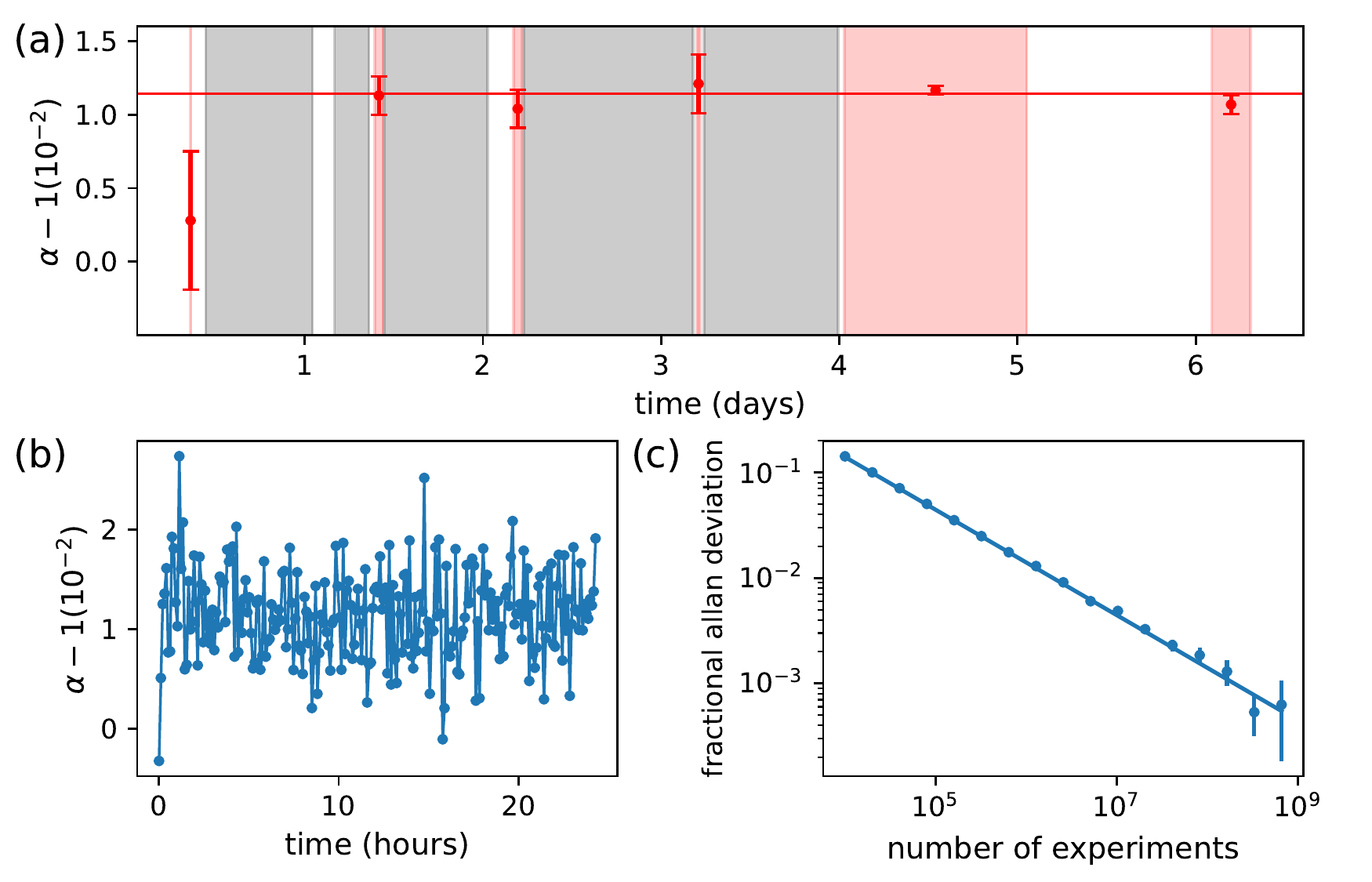}
\caption{Assessment of 650-nm background subtraction. (a) Red shaded regions indicate the time intervals during which $\alpha$ was assessed, with the red points indicating respective average results. Gray regions are the time intervals for the last five measurements given in \fref{fig:results}.  (b) From the 24 hour $\alpha$ measurement shown in (a), the average result per block of $10^7$ experiment cycles. (c) Fractional Allan deviation of (b). }
\label{fig:background}
\end{figure}

The five subsequent data sets were taken over four consecutive days (\fref{fig:background}(a), gray regions) with measurements of $\alpha$ interspersed.  After the final dataset, $\alpha$ was monitored over a full day and measured again two days later to check for stability and reproducibility.  Figure~\ref{fig:background}(a) indicates the measured $\alpha$ (red circles) along with the averaging time used in each case.   A plot of $\alpha$ from the longest measurement is shown in \fref{fig:background}(b-c) along with an Allan deviation of the data, which clearly shows it continues to average down throughout the day.  Note also that the smaller value of $\alpha$ from the first measurement is consistent with the shorter averaging time and the initial transient associated with the first two points in \fref{fig:background}(b).  This initial transient has a timescale on the order of 15\,minutes, which suggests a thermal response of the AOM.  Given the consistency of all measurements, and that data is taken over many hours, we take the weighted mean of all measurements, $\alpha=1.01143(27)$, to determine a correction.

From these considerations, we estimate fractional systematic shifts of approximately $4\times10^{-4}$ for each of the five datasets which are listed in \tref{tab:shifts}.  Although the stability and reproducibility of the background correction explains the consistency of all measurements, only for the last five data sets has this effect been assessed.  Hence only these data sets are included in the final analysis.

\subsection{Detector dead time}
\label{DeadTime}
Single photon counting modules have a dead-time, $\tau_d$, for which the detector is non-responsive to further photon arrivals subsequent to a detection event.  This reduces the count rate dependent on the photon arrival statistics and can therefore distort the measurement.  In addition there is a small probability of a secondary pulse being generated from a previous detection event; an effect known as after-pulsing.  This primarily provides a fractional increase in the mean counts and cancels in determining the branching ratio.  In principle, after-pulsing would have a secondary influence on dead-time effects but this would be a small correction to that determined from the dead time alone.  Since there is generally a low probability of two photons anyway, it can be anticipated that the effect of dead time is small and hence after-pulsing effects are neglected.

The largest systematic to the branching ratio reported in \cite{munshi2015precision} is due to detector dead time.  As with other reports \cite{ramm2013precision, fan2019measurement}, very little information was given on how this effect was assessed and quantified.  Here we give a detailed account on how the dead time is measured and how the systematic shift is assessed.  In addition, we give results of a numerical simulation which indicates that the assessment in \cite{munshi2015precision} is likely in error and largely explains the difference in the reported branching ratio.  The discussion given applies equally to either case of collecting photons from the $D_{3/2}$ to $P_{1/2}$ or $S_{1/2}$ to $P_{1/2}$ transitions so long as the branching ratio, $p$, refers to the probability of decay by the transition from which photons are collected.

\subsubsection{Measurement of the dead time.}
\label{Deadtime:Measure}
There are two relatively simple ways in which to measure a detector's dead time.  The first approach, as stated in \cite{munshi2015precision}, is to measure the SPCM count rate against a linear, actively stabilized light source.  The measured count-rate, $\gamma$ is given by~\cite{cantor1975dead}
\begin{equation}
\label{Eq:nonlinear}
\gamma=\frac{\gamma_i}{1+\gamma_i \tau_d}=\frac{\beta P}{1+\beta P \tau_d},
\end{equation}
where $\gamma_i$ is the expected count rate for a dead-time-free detector, and $\beta$ is an attenuation factor between the SPCM and the calibrated input power $P$.  So long as the detector is linear in the power, the nonlinear response to the input power is a measure of $\tau_d$.  The second approach is to use high resolution time-tagging to directly measure the arrival time statistics, with the dead time determined by the minimum arrival time seen in the data.

\begin{figure}
\includegraphics[width=\columnwidth]{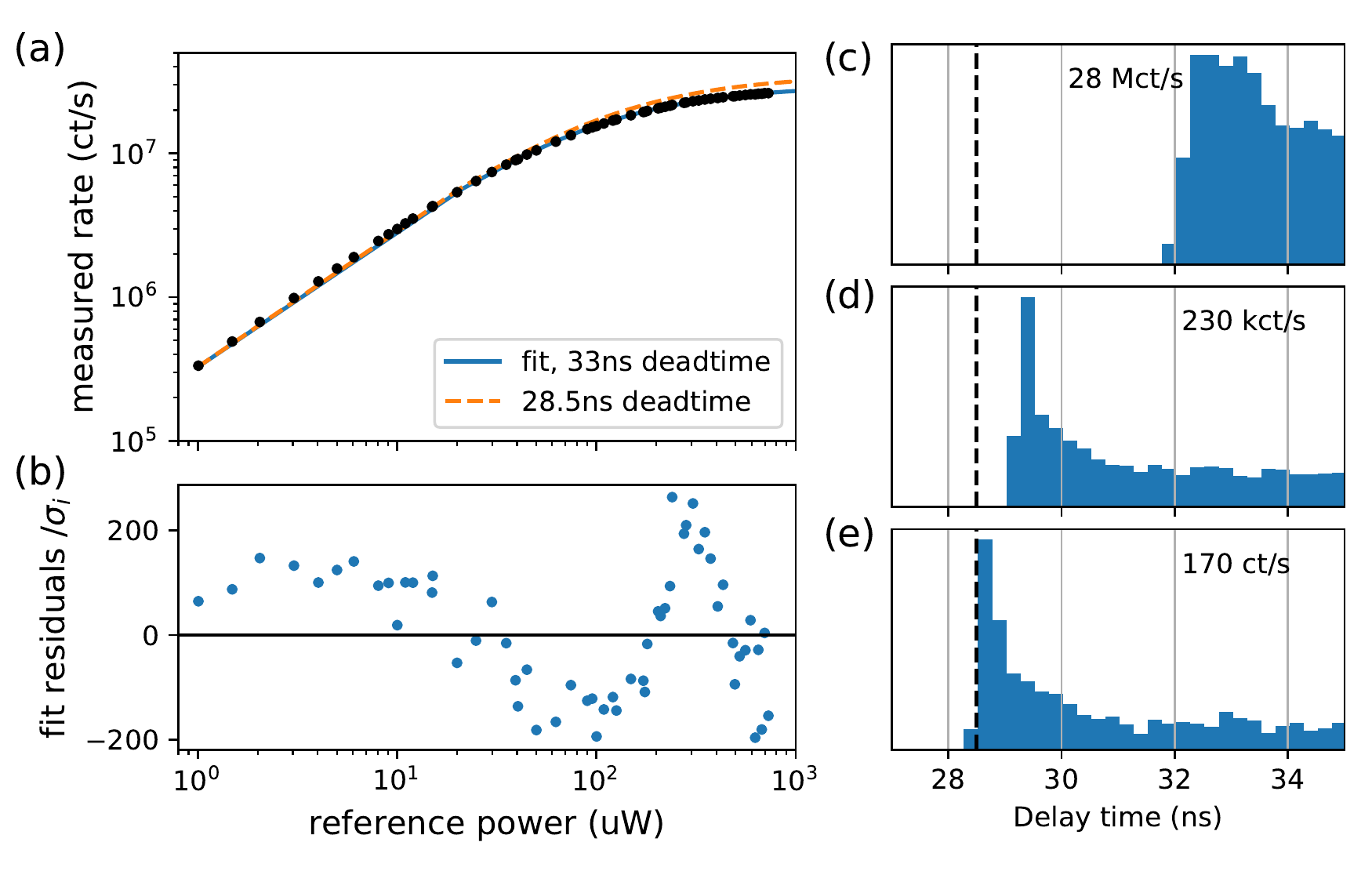}
\caption{Measurement of detector dead time: (a) The SPCM count rate is measured against a linear, actively stabilised light source. A dead time of 33 ns is extracted by fitting (blue line), but the residuals in (b) suggest a poor fit with Eq.~\ref{Eq:nonlinear}; (c)-(e) are arrival time statistics measured by high resolution (250\,ps) time tagging using different count rates as indicated in the plots. The detector model is SPCM-AQR-16 from Perkin-Elmer.}
\label{fig:deadtime}
\end{figure}

We have used both approaches with the results shown in Fig.~\ref{fig:deadtime}.  The SPCM count rate measured against a linear, actively stabilised light source is shown in Fig.~\ref{fig:deadtime}(a) from which a dead time of 33 ns is extracted. However the fitted residuals shown in Fig.~\ref{fig:deadtime}(b) suggest a poor fit with the model with an obvious correlation with input power. Later, the dead time was directly measured by high resolution (250 ps) time tagging. The detector dead time was found to vary from 28.5 ns at low saturation, to $\sim33$ ns at deep saturation, consistent with the fit in (a).  For the experiments here, the dead time at low saturation is relevant.  Due to the low count rates involved, all the time-tagged events shown in Fig.~\ref{fig:deadtime}(c) can be attributed to after-pulsing from which we deduce an after-pulse probability of $0.3\%$. The dead time, after-pulsing probability and tabulated linearity correction factor supplied by the manufacturer are all consistent with the measurements derived from the data shown in Fig.~\ref{fig:deadtime}.

An additional effect that occurs with SPCM's is twilighting: photons arriving near the end of a dead time may still be detected with the output pulse time-delayed to after the dead time. Thus a measured dead time may well be an over-estimate of an effective dead time relevant to an experiment. Measurements using a correlated photon source~\cite{stipvcevic2016advanced} and the same detector model series as used here suggest the effective dead time may be shorter: with events at 25\,ns having a detection efficiency of $\sim50\%$ relative to the maximum.  In light of our measurements, and the possible influence of twilighting, we use a dead time of 28.5\,ns to estimate the shift and allow a -2\,ns deviation as a systematic uncertainty.

\subsubsection{Dead time correction factor}
\label{Deadtime:Correction}
Determining a dead-time correction requires an estimate for the probability that two photons arrive within the detector dead time.  In the experiments reported here, background count rates are sufficiently low that they can be neglected.  Additionally, as the signal is at most one photon when pumping to $D_{3/2}$, dead time can only be significant when pumping back to $S_{1/2}$.  

\begin{figure}
\includegraphics[width=\columnwidth]{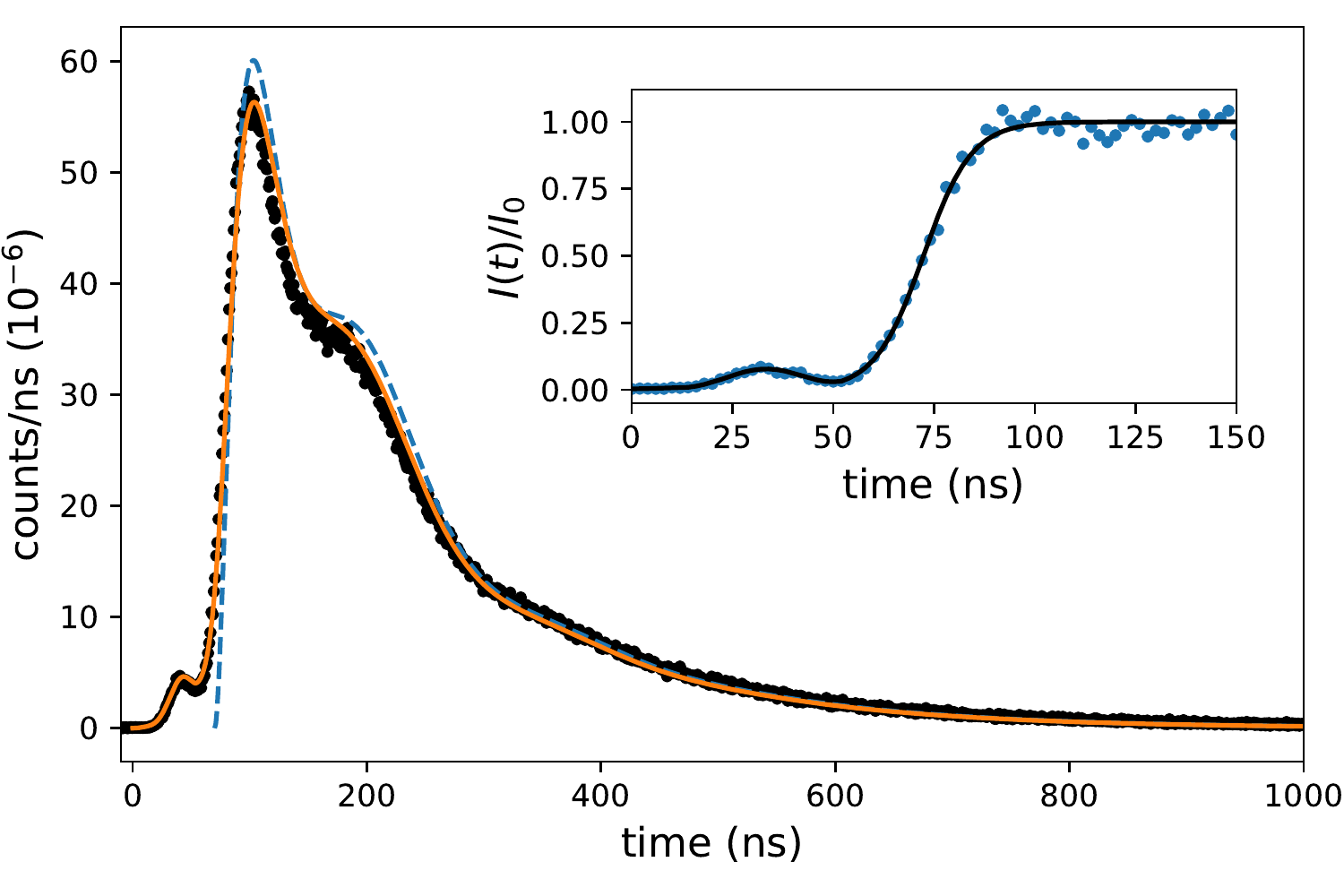}
\caption{Average number of counts measured (black dots) per 1 ns bin when pumping out of the $D_{3/2}$ for typical experimental parameters: magnetic field of $\sim0.22$ mT aligned to $\vop{x}$, 650 nm laser polarisation aligned to $\vop{y}$, and a pumping time of 152(6) ns into the $S_{1/2}$ state as measured in an independent experiment.  The blue line is the expected count rate obtained by integration of the master equation (given $p=0.268$ and $q=0.0268$) for a constant 650 laser coupling strength, which is chosen to yield the measured 152 ns pumping time into $S_{1/2}$.  The curve is offset by $70\,\mathrm{ns}$ for visual comparison.  The orange line is obtained by including a time-dependent coupling strength due to the measured switch on transient of the 650 laser as shown in the inset.}
\label{fig:reddist}
\end{figure}

An estimate of the dead time shift can be made by applying Eq.~\ref{Eq:nonlinear} to the dead-time-free rate predicted using a master equation to find $\gamma(t) = q p \Gamma \rho_{ee}(t)$ where $\rho_{ee}(t)$ and $\Gamma$ are the population and linewidth of $P_{1/2}$, respectively.  The parameters of the master equation are set as per the experiment, with the resonant laser coupling set such that the pumping time to $S_{1/2}$ matches the measured value.  Figure~\ref{fig:reddist} shows the average measured count rate, $\gamma(t)$, at a 1 ns resolution when pumping from $D_{3/2}$ to $S_{1/2}$ for a typical set of experiment parameters, which are given in the caption. For all configurations explored, we observe good agreement between the model and measured distribution with no free parameters. Strictly, one expects such a rate correction approach to be valid only when (i) the photon arrival times follow an exponential distribution and (ii) the rate dynamics are not faster than either the detector dead time or binning time. For the range of decay times and experiment configurations used here, this method implies shifts in the measured $p$ are $\sim1-2\times10^{-4}$.  Inclusion of the transient switch on time of the coupling (orange versus blue curve in \fref{fig:reddist}) reduces the shift by at most 6.7\% over all datasets.

While this approach may be justifiable for an ensemble of atoms and sufficiently slow dynamics, it is not for the transient emission from a single atom.  Given that the $P_{1/2}$ lifetime of $\sim 8\,\mathrm{ns}$ is a reasonable fraction of the dead time, the quantum nature of the emission process should also be considered. For a given branching ratio, $p$, and collection efficiency, $q$, this can be analysed by first considering the photon arrival statistics at a dead-time-free detector, and then determining how much the signal would be reduced by a dead time.

The distribution of counts for a dead-time-free detector are determined by a combination of a geometric distribution for the photon emissions, and a binomial distribution for the detection.  The net distribution for $k$ detected photons is then
\begin{equation}
f(k;p,q)=\frac{(1-p)p^k q^k}{(1-p(1-q))^{k+1}}.
\end{equation}
A dead time will change the distribution and reduce the mean counts but, in principle, the mean can be still expressed in the similar form
\begin{equation}
\label{mod_r_Eq}
r=\sum_k f(k,p,q) \bar{k}_k,
\end{equation}
where $\bar{k}_k$ is an effective mean number of counts given $k$ counts in the ideal case.  Expressing the mean in this way provides a viable approach for determining the dead time correction: once $\bar{k}_k$ is determined,  an estimate, $\bar{p}$, can be computed from Eq.~\ref{branchEq} using the modified expression for $r$ given in Eq.~\ref{mod_r_Eq}, and a correction estimated by the difference $p-\bar{p}$.  Experimentally, $\bar{p}$ is similarly estimated from a measured $r$, and it can be readily verified that $\bar{k}_k=k$ gives $\bar{p}=p$.  Note that $p$ is set to a fixed value.  Since the correction $p-\bar{p}$ is insensitive to small changes in $p$ it is sufficient to set it to $0.268$ in estimating the correction.

In general $\bar{k}_k$ is complicated to compute but it is tractable for $k=2$ and $k>2$ terms do not significantly contribute for the low count rates involved.  For the $k=2$ case, one can compute the probability of having two photons with $m$ photons missed in between and show that $\bar{k}_2$ is given by
\begin{equation}
\bar{k}_2=2-(1-p(1-q)) \sum_{m=0}^\infty p^m(1-q)^m p_m,
\end{equation}
where 
\begin{equation}
p_m=\int_0^{\tau_d} P_m(t)dt,
\end{equation}
and $P_m(t)$ are the recursively defined functions
\begin{equation}
P_m(t)=\int_0^t P_{m-1}(t-x) P_0(x)dx,\quad P_0(t)=P(t),
\end{equation}
giving the arrival time probability distribution of a subsequent detected photon, given that $m$ photons were missed.  The function $P(t)$ is the probability distribution for single photon emission times and can be found from a master equation formulation as outlined in the appendix.

Although the $k=3$ term does not significantly contribute, it is useful to have a lower bound for its contribution.  This can be obtained by neglecting any missed events and determining the three mutually exclusive possibilities: all photons arrive within the dead-time of the first (1 count), all photons are separated by at least the dead-time (3 counts), and everything in between (2 counts).  This will over-estimate $\bar{k}_3$, as missed events increase the time separation of detected photons, but it provides a bound on its contribution to the dead-time correction.  With $p_{3,k}$ denoting the probability that $3$ photons give $k$ counts, we have
\begin{equation}
p_{3,1}=\int_0^{\tau_d} \int_0^{\tau_d-t_2} P(t_2)P(t_1)dt_1 dt_2,
\end{equation}
$p_{3,3}=(1-p_0)^2$, and $p_{3,2}=1-p_{3,3}-p_{3,1}$.  The mean is then readily computed giving
\begin{equation}
\tilde{k}_3=\sum_{k=1}^3 k p_{3,k} \lesssim \bar{k}_3 < 3.
\end{equation}

From this procedure, the estimated dead time corrections to $p$ are $\lesssim 2\times 10^{-4}$ with an uncertainty from $\bar{k}_3$ being only $1-2\%$ of this.  However, there are two effects that can lead to a systematic over-estimation of the correction.  The first, as already mentioned, is due to twilighting, which can reduce the effective dead time from the measured value.  From our analysis, a 2\,ns reduction in the effective dead time reduces the correction by $\sim13\%$.  The second, is due to the temporal profile of the pulse, more specifically the turn on time.  The analysis above assumed a constant laser coupling, which is not strictly the case.  As a variable coupling makes the function $P(t)$ dependent on the emission time of the previous photon, it is not easy to incorporate this into the analysis.  However, as the correction derived by our treatment and that from a rate correction approach agree to within $20\%$ for all datasets, we use the latter to estimate this systematic.  That is, for each dataset the fractional systematic effect of the transient on the dead time correction is estimated using the rate correction method and we add this correction to the systematic error from the effective dead time uncertainty.

\subsubsection{Numerical experiments}
A simulation was developed that generated random detection times for a given experiment.  Since detection times between different atoms and background counts are uncorrelated, multiple atoms and background effects could be included by simply merging datasets.  The simulation assumed no dead time and dead-time effects assessed as a post processing step that eliminated counts that appeared within a given dead time of a prior event.  In essence, the simulation consisted of the following steps
\begin{enumerate}
\item Generate $N$ random numbers, $n$, from a geometric distribution characterized by $p$.  This gives the number of red photon emissions for a particular experiment.
\item For each $n$, generate emission times from the distribution $P(t)$ found from the master equation and select from them based on Bernoulli trials with probability $q$.  This represents the detected arrivals at the detector. 
\end{enumerate}
Each class of $n$ emissions could be treated vectorially allowing efficient simulation within Python.  For the single atom case reported here, up to a billion experiments could be simulated in $\sim100\,\mathrm{s}$.  For the two extreme pumping times used in the experiments, we were able to confirm our analysis at the $10^{-5}$ level.  In addition, a preliminary simulation of the experiment in \cite{munshi2015precision} suggests this is a likely source of the discrepancy.  Using pumping times consistent with their published data with a number of ions and collection efficiency consistent with their signal strengths, we estimate a dead-time correction factor of $\sim6\times10^{-3}$ compared to their reported $4\times10^{-3}$.  This would increase their reported branching ratio for $P_{1/2}$ to $S_{1/2}$ from 0.7293 to 0.7313, and reduce the corresponding value for $P_{1/2}$ to $D_{3/2}$ to 0.2687.  Moreover, we see variations in the correction factor of $\lesssim 10^{-3}$ for differing atom number and pumping rates. 

\subsection{Detection Optics}
\label{Detection}
For a decay from $J'=1/2$ to either $J=1/2$ or $J=3/2$ there is a 1/3 chance that the atom will decay by a $\pi$ transition, independent of which upper Zeeman state is involved.  Given that the intensity distributions for $\sigma^\pm$ emissions are the same, the emission from $J'=1/2$ to either $J=1/2$ of $J=3/2$ is always isotropic.  However, as noted in other reports \cite{munshi2015precision,ramm2013precision,fan2019measurement}, the polarisation may well depend on the B-field orientation, as illustrated by the Hanle effect.  Consequently, polarisation sensitivity of the collection optics could give rise to an orientation and excitation dependence to the photon collection that could potentially bias the measurement.  To address this, the common procedure has been to repeat the measurements at different field orientations to test for such an effect.  As the effect depends on the laser polarization, orientation of the magnetic field, and imaging direction, it is important to clarify the configurations used.

The optical imaging system consists of a back-to-back lens doublet with the imaging axis normal to the vacuum window.  The alignment is optimised by minimising spherical aberrations apparent in the de-focussed image of a single ion and ensuring these have spherical symmetry.  This ensures the optical axis is normal to the window and the ion centred on the imaging axis.  In this configuration any birefringence would unlikely result in a polarisation selectivity.   Measurements with an independent laser verify this to the $\lesssim1\%$ level limited by the power stability of the beam.  Although this is not enough to rule out a problem, the beam configuration can further mitigate the effects of any such problem. 

As described earlier, the pumping lasers propagate along $\vop{z}$, which is orthogonal to the imaging axis ($\vop{x}$) and at 45 degrees to the trap axis.  Both beams are linearly polarised with the 493-nm and 650-nm lasers polarised along $\vop{x}$ and $\vop{y}$, respectively.  In all experiments this configuration is unchanged.  Linear polarisation of the excitation lasers precludes any magnetic field dependence measured by polarisation sensitive optics.

With the magnetic field aligned along $\vop{z}$, both pump beams can only excite to a single upper $m$ state and polarisation components will not coherently interfere, even if the light field polarisations were slightly elliptical.  Even if there was an imbalance in the $\sigma^\pm$ emissions, both project along $\vop{y}$ when propagating along $\vop{x}$.  Moreover, when propagating in the half space $z>0$ ($z<0$), $\sigma^+$ would project preferentially to right (left) circular polarisation and vice-versa for $\sigma^-$.  Thus, the imaging optics could not distinguish the two components unless the optics also had a spatial dependence to any already unlikely polarisation selectivity.

It might be argued that the alignments will not be perfect which may result in coherence effects between $\vop{x}$ and $\vop{y}$ components of the outwardly propagating light field.  This would require a circular component to the laser polarisation with propagation oblique to the magnetic field.  To make this manifestly more prominent, measurements were made with the magnetic field aligned along $(\vop{y}+\vop{z})/\sqrt{2}$, noting that it cannot be aligned along $\vop{y}$ without diminishing the efficacy of optical pumping with the 650-nm laser for the polarisation used; along $\vop{y}$ would be a textbook Hanle effect, at least for circular polarisation of the excitation beam.  

Measurements were also made with the magnetic field aligned along $\vop{x}$ in which case only $\sigma^\pm$ components are detected.  For reasons not related to this experiment, measurements were also made with the field at $\sim 30^\circ$ to the imaging axis and in the horizontal plane $\sim(\sqrt{6} \vop{x}+\vop{y}+\vop{z})/\sqrt{8}$.  

Given that there is no statistically significant variation in the measurements, and no compelling reason to believe there should be an orientation dependence to the results given the set up used, we conclude that there is no systematic error associated with any polarisation sensitivity of the detection optics; hence no error or uncertainty is attributed to this effect.

\subsection{Miscellaneous systematics}
\label{Misc}
\subsubsection{Lifetime of the $D_{3/2}$ level}
During the optical pumping step from $S_{1/2}$ to $D_{3/2}$, there is a small probability that decay from $D_{3/2}$ repopulates $S_{1/2}$ resulting in an increase in the signal measured during this step.  Inasmuch as the pumping to $D_{3/2}$ can be described by an exponential with time scale $\tau_b$, which is much smaller that duration of the optical pumping, $T$, the probability of decay is given by
\begin{equation}
\gamma \int_0^T P_{D_{3/2}}\,dt \approx \gamma \int_0^T 1-e^{-t/\tau_b}\,dt\approx\gamma (T-\tau_b),
\end{equation}
where $P_{D_{3/2}}$ denotes the population in $D_{3/2}$ and $\gamma$ its decay rate.  As decay from $D_{3/2}$ results in repumping, this probability then represents the fractional increase in the measured signal due to the finite lifetime.  However, the background is similarly affected by the decay.  In this case the system starts in this state and the decay probability becomes $\gamma T$.  Hence the fractional change in the background-subtracted signal is $-\gamma \tau_b$.

Optical pumping  from $D_{3/2}$ to $S_{1/2}$ is similarly affected.  However in this case the background is unaffected and $D_{3/2}$ level is only transiently occupied.  This gives a decay probability of approximately $\gamma \tau_r$, which is the fractional decrease in the measured signal.  To the extent that both optical pumping rates are equal, there is therefore a cancellation in the ratio of the two signals.  In the case of Ba$^+$ considered here, $\gamma\sim 0.0125/\mathrm{s}$ and $\gamma \tau_{b,r}< 5\times 10^{-9}$.  Thus the effect is negligible, independent of the cancellation.

\subsubsection{Off-resonant excitation to the $P_{3/2}$ level.}
During optical pumping, there is a small probability of off-resonant excitation to $P_{3/2}$.  In this event, the atom can scatter to $S_{1/2}$, $D_{3/2}$, or $D_{5/2}$  with probability $p_{1/2}$, $p_{3/2}$ or $p_{5/2}$, respectively.  Scattering to $D_{5/2}$ places the ion in a dark state, which, in the case of Ba$^+$, persists for $\sim30\,\mathrm{s}$ and is readily detected in the experiment.  Moreover, events that scatter back into the same level are inconsequential.  Thus, when scattering from $S_{1/2}$ to $D_{3/2}$ the fractional decrease in the signal is determined by
\begin{equation}
p_{3/2} \Gamma' \frac{\Omega_b'^2}{4\Delta^2}\tau_b,
\end{equation}
where $\Gamma'$ is the linewidth of the $P_{3/2}$ level, $\Omega_b'$ is the coupling strength of the 493 laser to $P_{3/2}$, and $\Delta$ is approximately the fine-structure splitting.  A similar expression holds when optically pumping from $D_{3/2}$ to $S_{1/2}$.  Owing to rapid pumping times ($\tau_{b,r}\leq 300\,\mathrm{ns}$) and large fine structure splitting $\Delta=2\pi\times 50.7\,\mathrm{THz}$, this effect is negligible even under extreme circumstances.

Although direct off-resonant excitation can play no significant role in the branching ratio measurement, it may well happen that the pump lasers have a broadband pedestal, which could significantly increase the scattering rate if the pedestal is near resonant.  In the case of the Ba$^+$ experiment reported here this cannot occur: the 493-nm laser driving the $S_{1/2}$ to $P_{1/2}$ is frequency doubled, heavily suppressing any such pedestal, and the 650-nm laser driving the $D_{3/2}$ to $P_{1/2}$ transition has no significant gain at the required resonant wavelength of 585\,nm.  Moreover the optics used in the experiment do not support this wavelength.

\subsubsection{AOM extinction}
The finite extinction ratio of the AOMs used to switch the pumping beams provides an effective decay rate from one state to the other and can be treated in a similar manner as a finite state lifetime.  When optically pumping from $S_{1/2}$ to $D_{3/2}$, residual light at 650\,nm provides an effective decay of the $D_{3/2}$ level given by $\gamma_r=\alpha_r/\tau_r$, where $\alpha_r$ is the extinction ratio of the AOM switch.  As in the case of a finite lifetime, this results in a fractional shift $-\alpha_r\tau_b/\tau_r$ of the background-subtracted signal.  Similarly, when optical pumping from $D_{3/2}$ to $S_{1/2}$, there is a fractional shift of $-\alpha_b\tau_r/\tau_b$.  

For the current implementation in Ba$^+$, measured extinction ratios for both beams are better than $5\times 10^{-7}$.  Moreover the majority of this light is unshifted in frequency and hence far detuned from the respective transitions.  Even neglecting this frequency dependence, the effect on the measured branching ratio is then $\lesssim 10^{-7}$.
  
\subsubsection{Finite pumping times}
The finite duration of pumping times can result in either a loss of accuracy or a loss in precision.  If the pumping times are set too long, it takes too long to accumulate enough data to reach a desired precision.  If the pumping time is too short, the optical pumping is incomplete resulting in a systematic shift of the measured signals.  In general, if the population left behind during an optical pumping step is $\epsilon\ll 1$, this population will be transferred during the second background pulse.  Hence the net change in the background-subtracted signal is $2\epsilon$. In the experiments reported here, the optical pumping times are at least $20\tau$ where $\tau$ is the measured decay rate for the signal of interest.    Even allowing for a non-exponential decay that arises in such multi-level pumping, systematic shifts associated with incomplete pumping are well below any realistically achievable statistical uncertainty.

\section{Summary}
We have provided measurements of the branching ratio for $6P_{1/2}$ decays to $5D_{3/2}$ in $^{138}$Ba$^+$ obtaining a value of $p=0.268177\pm(37)_\mathrm{stat}-(20)_\mathrm{sys}$.  Together with the experimentally determined reduced electric dipole matrix element $\langle{6 P_{1/2}}\|r\|{6 S_{1/2}}\rangle=3.3251(21)\,\mathrm{a.u.}$ reported in \cite{woods2010dipole}, we obtain an estimate of $\tau=7.855(10)\,\mathrm{ns}$ for the $P_{1/2}$ lifetime and $3.0413(21)\,\mathrm{a.u.}$ for the matrix element $\langle{6 P_{1/2}}\|r\|{5 D_{3/2}}\rangle$.  From theoretical considerations, we also extract the matrix elements, $\langle{6 P_{3/2}}\|r\|{5 D_{3/2}}\rangle=1.3285(13)\,\mathrm{a.u.}$ and $\langle{6 P_{3/2}}\|r\|{5 D_{5/2}}\rangle=4.091(3)\,\mathrm{a.u.}$ The latter matrix element is a prominent contributor to $\Delta\alpha_0(\omega)$ of the $S_{1/2}$-to-$D_{5/2}$ clock transition and will allow a precise model of $\Delta\alpha_0(\omega)$ to be constructed as discussed in \cite{BarrettProposal}.  The matrix elements also provide a new estimate of $6.271(8)\,\mathrm{ns}$ for the $P_{3/2}$ lifetime. 

In the course of this work, we have identified a systematic that has not been considered in previous reports and arises from imperfect background subtraction. It might be argued that, in this work, photons are collected on the transition with the smallest branching ratio and from a single atom, such that this effect is only important here.  Although it is true that the signal is indeed lower here than in other works, what is important is the SNR, with the shift given by 
\begin{equation}
\Delta p=\frac{(1-p)p}{\mathrm{SNR}}(\alpha-1),
\end{equation}
and the uncertainty determined by replacing $(\alpha-1)$ with $\delta\alpha$.  The typical background seen in Fig.~\ref{fig:timing} is significantly below that seen in \cite{ramm2013precision,munshi2015precision, likforman2016precision, zhang2016iterative}.  Indeed, the background here is only visible in Fig.~\ref{fig:timing} due to the use of a log scale.  Estimates from the respective figures in \cite{ramm2013precision,munshi2015precision} would suggest a lower SNR in those cases in spite of using a transition producing more photons that are collected from multiple atoms.  With a SNR of 2 \cite{likforman2016precision} or even less \cite{munshi2015precision, zhang2016iterative}, the efficacy of background subtraction cannot be simply assumed.

In \cite{ramm2013precision}, it is suggested that statistical error could be improved in this measurement scheme by increasing detection efficiency.  In general, we would disagree.  As with any experiment, accuracy and precision is limited by SNR, equipment calibration, and the number of measurements made.  In this measurement scheme, SNR manifests in background subtraction, equipment calibration comes down to detector dead-time corrections, and, with due consideration to photon statistics, the statistical accuracy is
\begin{equation}
\delta p =\frac{(1-p)p}{\sqrt{Npq}}.
\end{equation}
As explicitly shown here, background subtraction would likely average down such that it would not be a limitation; although clearly it must still be calibrated.  Given that the dead time has a complex dependence on the detector implementation \cite{stipvcevic2016advanced}, and the shift it induces is dependent on the excitation scheme, it would be difficult to calibrate a large shift to high accuracy.  From statistical considerations, the experiment cycle time would be made as fast possible until dead time effects become important.  At that point, improving detection efficiency, $q$, using the transition with higher value of $p$, or increasing the number of atoms would only make it more difficult to calibrate a dead-time effect.

In \cite{ramm2013precision}, two pumping pulses of different intensity were used to reduce dead-time effects.   More generally, for a given experiment duration one could consider arbitrary temporal shaping of the pulse to minimize dead-time effects without compromising the finite pumping error and SNR.   Simulated dynamics for a linearly ramped pulse intensity, where the dead-time correction is estimated by \eref{Eq:nonlinear}, indicate this strategy can reduce dead-time effects and may have a role in experiment optimisation.  However, proper evaluation of the dead-time shift would require careful consideration and a model beyond that presented in \sref{Deadtime:Correction}.

In all measurements of this sort, the potential for a Hanle type of effect to bias the result is typically explored by changing the orientation of the field.  We have also done this, along with qualification of those orientations used.  In retrospect, a better way to investigate this potential systematic would be to explicitly use a configuration in which the Hanle effect should be observable if the detection was polarisation sensitive i.e having circularly polarised light for both beams, and the magnetic field along $\hat{\mathbf{y}}$ with sufficient amplitude to maximise a potential discrepancy, which we estimate to be $0.06$ in the measured branching ratio for our case.  A consistent branching ratio measurement would then bound the polarisation selectivity of the optics and allow more stringent estimates on the effect for configurations inherently insensitive to the effect.

The result reported here also illustrates the importance of reproducing results within the scientific literature.  Results such as these can be factored into other experimental measurements or used as benchmarks against theoretical calculations.  Given the combination of measurements and theory used here to provide new estimates of matrix elements, lifetimes, and branching ratios, it would be of interest to directly measure matrix elements by a different methodology such as that demonstrated in \cite{hettrich2015measurement, arnold2019dynamic}.  This would provide stringent tests of multiple precision measurements and theory.

\begin{acknowledgements}
This work is supported by the National Research Foundation, Prime Ministers Office, Singapore and the Ministry of Education, Singapore under the Research Centres of Excellence programme. This work is also supported by A*STAR SERC 2015 Public Sector Research Funding (PSF) Grant (SERC Project No: 1521200080). T. R. Tan acknowledges support from the Lee Kuan Yew post-doctoral fellowship. M. S. S. acknowledges support from the Office of Naval Research, Grant No. N00014-17-1-2252.
\end{acknowledgements}

\appendix
\section{Photon emission distribution}
The distribution function $P(t)$ can be determined from the master equation
\begin{equation*}
\dot{\rho}=-i[H,\rho]-\frac{\Gamma}{2}\left(P_e \rho+\rho P_e\right)+\sum_{\substack{\alpha=r,b\\ \lambda=0,\pm}}\gamma_\alpha\left(A_{\alpha,\lambda}\rho A^\dagger_{\alpha,\lambda}\right),
\end{equation*}
where $A_{\alpha,\lambda}$ is the dipole operator for the red and blue decay channels and polarisation $\hat{\mathbf{e}}_\lambda$, $\gamma_\alpha$ are the partial decay rates related to the total decay rate $\Gamma$ via $\gamma_r=p \Gamma$ and $\gamma_b=(1-p) \Gamma$ and 
\begin{equation}
P_e=\sum_\lambda A^\dagger_{\alpha,\lambda}A_{\alpha,\lambda}
\end{equation}
is the projection operator onto the excited states.  Integration for our eight level system is numerically straight-forward.  

For the Hamiltonian, we assume the 650-nm laser is on resonance and the magnetic field set as per the experiment. We then integrate the full equations for a given branching ratio and laser coupling strength with an initial distribution that uniformly occupies all sub-levels of $D_{3/2}$.  The branching ratio is set to the estimated value $0.268$ and the laser coupling adjusted to give a decay rate of the $D_{3/2}$ population that matches that observed in the experiment. 

Once the laser coupling is set, the distribution function, $P(t)$ can be found by integrating the equations with the final term in the master equation omitted.  The desired function $P(t)$ is then given by
\begin{equation}
P(t)=\Gamma \mathrm{Tr}(P_e \rho(t)),
\end{equation}
from which all the desired expressions can be calculated numerically.

Evidently, the function $P(t)$ has no explicit dependence on the branching ratio since the branching ratio only comes into the last term of the master equation, which is dropped from the integration.  This reflects the fact that the branching ratio determines the state to which atoms decays and not the fact that it actually decays in the first place.  However, it implicitly depends on the branching ratio based on how we set the Hamiltonian. 

\bibliography{BariumBranching}
\bibliographystyle{unsrt}

\end{document}